\documentclass[prb,amsmath,amssymb,superscriptaddress,groupaddress,
showpacs,twocolumn,reprint]{revtex4-1}

\usepackage{graphicx}
\usepackage[utf8]{inputenc}
\usepackage{color}
\usepackage{siunitx}
\usepackage{braket}
\usepackage{amsfonts}

\newcommand{\vsd}{\ensuremath{V_{\text{sd}}}}
\newcommand{\vg}{\ensuremath{V_{\text{gate}}}}
\newcommand{\didv}{\ensuremath{\text{d}I/\text{d}\vsd}}
\newcommand{\Tc}{\ensuremath{T_{\text{c}}}}
\newcommand{\Bc}{\ensuremath{B_{\text{c}}}}

\newcommand{\DSO}{\ensuremath{\Delta_\text{\tiny{SO}}}}
\newcommand{\DKK}{\ensuremath{\Delta_\text{\tiny{KK'}}}}

\newcommand{\dan}[1]{\operatorname{\hat d}_{#1}}
\newcommand{\dddag}[1]{\operatorname{\hat d}^{\dagger}_{#1}}

\newcommand{\Hop}[1]{\operatorname{\hat H_{#1}}}

\newcommand{\un}[1]{{\,\text{#1}}}

\begin{document}

\title{Sub-gap spectroscopy of thermally excited quasiparticles in a Nb 
contacted carbon nanotube quantum dot}

\author{M.~Gaass}
\affiliation{Institute for Exp.~and Applied
Physics, University of Regensburg, 93040 Regensburg, Germany}

\author{S.~Pfaller}
\email{sebastian1.pfaller@ur.de}
\affiliation{Institute for Theoretical 
Physics, University of Regensburg, 93040 Regensburg, Germany}

\author{T.~Geiger}
\affiliation{Institute for Exp.~and Applied
Physics, University of Regensburg, 93040 Regensburg, Germany}

\author{A.~Donarini}
\affiliation{Institute for Theoretical 
Physics, University of Regensburg, 93040 Regensburg, Germany}

\author{M.~Grifoni}
\affiliation{Institute for Theoretical 
Physics, University of Regensburg, 93040 Regensburg, Germany}

\author{A.~K.~Hüttel}
\affiliation{Institute for Exp.~and Applied
Physics, University of Regensburg, 93040 Regensburg, Germany}

\author{Ch.~Strunk}
\email{christoph.strunk@ur.de}
\affiliation{Institute for Exp.~and Applied 
Physics, University of Regensburg, 93040 Regensburg, Germany}

\date{\today}


\begin{abstract}
We present electronic transport measurements of a single wall carbon nanotube 
quantum dot coupled to Nb superconducting contacts. For temperatures comparable 
to the superconducting gap peculiar transport features are observed 
inside the Coulomb blockade and superconducting energy gap regions. 
The observed temperature dependence can be explained in terms of sequential 
tunneling processes involving thermally excited quasiparticles.
In  particular, these new channels give rise to two unusual conductance peaks 
at zero bias in the vicinity of the charge degeneracy point and allow to 
determine the degeneracy of the ground states involved in transport. The 
measurements are in good agreement with model calculations.
\end{abstract}

\maketitle

{\it Introduction--} 
Carbon nanotubes (CNTs) are highly versatile quantum systems, whose properties 
can be investigated by attaching them to a wide variety of different contact 
materials.\cite{Bockrath1997a, prb-jensen, prb-kasumov} By using 
superconducting metals as electrodes, a significant increase of spectroscopic 
resolution due to the sharp peaks at the gap edges in the BCS density of states 
can be achieved.\cite{prb-groverasmussen} Depending on the coupling strength 
between the carbon nanotube and its leads, the nanotube can act as a 
Josephson weak link, 
and proximity-induced supercurrent can flow through the quantum 
dot.\cite{prb-kasumov, nnano-cleuziou, JarilloHerrero2006, Pallecchi2008} The 
supercurrent is carried by Andreev bound states, whose 
presence is revealed by peculiar subgap features.\cite{nphys-dirks, 
nphys-pillet, Kim2013, Pillet2013, Kumar2014} By fabricating the contacts from 
sputtered Nb, they can remain superconducting up to a critical 
temperature $\Tc=\SI{8.5}{\kelvin}$ and a correspondingly large critical 
magnetic field $\Bc$.

In this work we report on sub-gap features observed in a CNT quantum dot weakly 
coupled to superconducting leads. Strikingly, such features are not visible at 
the lowest temperatures achieved in the experiment but only when the 
temperature becomes comparable to the superconducting gap. 
This suggests that, as explained below, the observed sub-gap features are not
due to Andreev reflections 
but rather to thermal excitation of quasiparticles across the gap, as predicted
recently by some of us.\cite{Pfaller2013}
We perform a systematic analysis of the 
temperature dependence of the observed features. A good agreement between 
experimental data and theoretical predictions in the linear as well as in the 
nonlinear regime is obtained.

{\it Experimental details--} 
The measurements presented here were performed on a single wall carbon nanotube 
grown by chemical vapour deposition (CVD).\cite{Kong1998} As substrate highly 
p-doped Si capped with \SI{300}{\nano\meter} $\text{SiO}_\text{x}$ is used. The 
electrodes to the nanotube are composed of \SI{3}{\nano\meter} Pd as contact 
layer and \SI{45}{\nano\meter} sputtered Nb with a contact spacing of the order 
of \SI{300}{\nano\meter}. The room temperature resistance of our device 
is in the range of \SI{100}{\kilo\ohm}. 

For performing two- and four-point measurements, each superconducting 
electrode is connected to two AuPd leads as resistive on-chip elements that 
are, among other filter stages, supposed to damp oscillations at the plasma 
frequency of the Josephson junction.\cite{Martinis1989, Pallecchi2008} 
\begin{figure}[ht]
\includegraphics[width=\columnwidth]{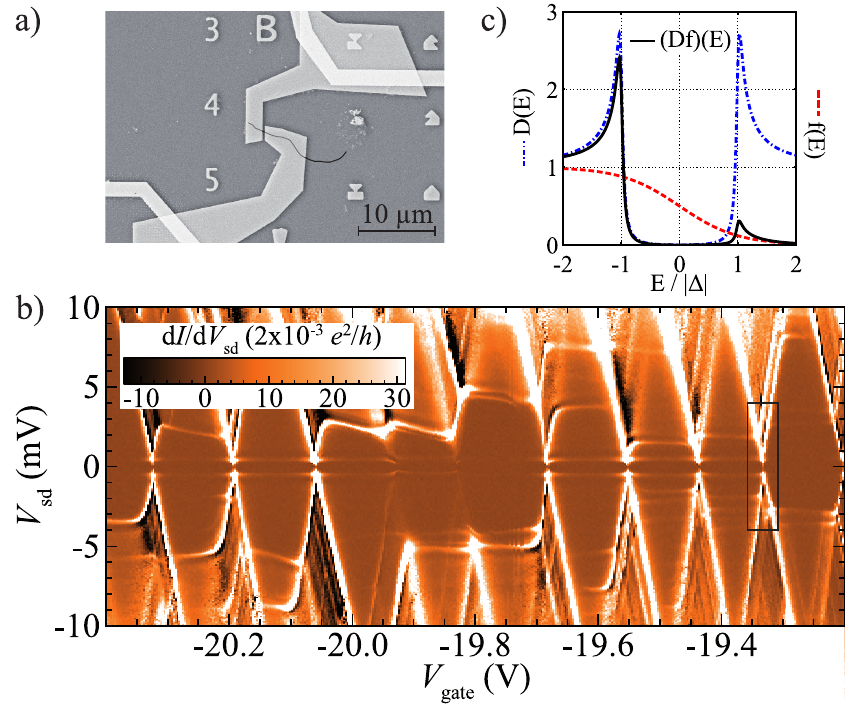}
\caption{
(Color online) (a) Scanning electron micrograph of the measured device 
displaying the resistive AuPd leads (bright) and the Nb electrodes (faint).
The location of the nanotube is indicated by the black curve. (b) Differential
conductance $\text{d}I(\vsd,\vg)/\text{d}\vsd$ as a function of source-drain
voltage \vsd\ and gate voltage \vg\ at $T= \SI{25}{\milli\kelvin}$. Dark areas
correspond to negative differential conductance. The black rectangle outlines
the parameter region of Fig.~\ref{Fig:Figure2}. 
(c) Scheme explaining the thermal excitation of quasiparticles across the
superconducting gap, see text.
}
\label{Fig:Figure1}
\end{figure}
A scanning electron micrograph of the sample is shown in 
Fig.~\ref{Fig:Figure1}(a). The device was measured in a dilution refrigerator 
with a base temperature of \SI{25}{\milli\kelvin}.

{\it Transport spectroscopy--} 
Fig.~\ref{Fig:Figure1}(b) shows an overview plot of the differential 
conductance \didv\ as a function of source-drain voltage \vsd\ and gate 
voltage \vg\ at $T=\SI{25}{\milli\kelvin}$. This temperature is much 
smaller than the critical temperature $T_c$ expected for our Nb contacts. The 
measurement of Fig.~\ref{Fig:Figure1}(b) serves as a reference for the high 
temperature experiments and theoretical predictions discussed below. Besides 
regular Coulomb diamonds, a rich substructure of both elastic and inelastic 
cotunneling lines is observed,\cite{DeFranceschi2001, prb-groverasmussen, 
Holm2008} reflecting the high spectroscopic resolution brought about by the 
sharp peaks in the BCS density of states (cf. Fig.~\ref{Fig:Figure1}(c)). 

The superconducting energy gap estimated from the sequential tunneling features 
at \mbox{$\vsd = \pm {2\Delta}/{e}$ } (see details below) is \mbox{$\Delta \sim 
\SI{320}{\micro\electronvolt}$}, compared to 
an expected value of \makebox{$\Delta = \SI{1.5}{\milli\electronvolt}$} for 
bulk Nb.\footnote{An evaluation of the elastic cotunneling lines in 
Fig.~\ref{Fig:Figure1}(b), not within the scope of our lowest-order theory, 
results in a slightly reduced value $\Delta \sim 
\SI{250}{\micro\electronvolt}$.} This reduction of the 
gap has been reported before in similar Nb-based 
devices.\cite{prb-groverasmussen,Kumar2014} Its origin so far remains an open 
question, though contamination of the lower Nb interface, formation of niobium 
oxide\cite{Hulm1972}, or the thin Pd contact layer may play a 
role.\cite{prb-groverasmussen} 
Estimated from \makebox{$\Delta = \SI{320}{\micro\electronvolt}$}, the 
resulting effective critical temperature would be \makebox{$\Tc \sim 
\SI{2.1}{\kelvin}$}. However, features in the data attributable to
superconductivity remain present up to temperatures of about 
\SIrange{3}{5}{\kelvin}, and measurements of a co-deposited Nb strip of 
comparable dimensions on the same chip yielded a critical temperature of 
\makebox{$\Tc = \SI{8.5}{\kelvin}$}.

From additional stability diagrams similar to 
Fig.~\ref{Fig:Figure1}(b) but taken at higher temperatures and finite magnetic 
field to suppress superconductivity (not shown), we estimate a charging energy 
$U \sim \SI{15}{\milli\electronvolt}$. From the fitting between 
experiments and theory discussed below (cf. in particular 
Eq.~(\ref{eq:conductance})), a coupling strength between quantum dot and 
leads of \makebox{$\Gamma \sim \SI{0.093}{\milli\electronvolt}$} is
extracted.
 This places our measurement into the parameter range  \makebox{$ 
\Gamma < \Delta \ll U$} where Coulomb repulsion dominates transport, 
superconductivity enhances the spectroscopic
resolution, \cite{DeFranceschi2010} and Andreev reflections are
expected to 
be strongly suppressed.\cite{Buitelaar2003} 
No obvious traces of Kondo phenomena 
\cite{GoldhaberPRL1998} are observed neither in the normal nor in the 
superconducting state.

{\it Thermally activated transport--}
For quantum dots connected to superconducting leads, transport is usually 
blocked in the energy gap range $\left| e\vsd\right| \le 2 \Delta$. At high 
temperature, transport becomes possible both at low bias and in parts of the 
Coulomb blockade region due to quasiparticles excited across the superconducting 
energy gap.\cite{Pfaller2013} This 
is illustrated in Fig.~\ref{Fig:Figure1}(c), showing the product (black solid 
line) of the quasiparticle density of states (blue dash-dotted line) and the 
Fermi function (red dotted line). For sufficiently high temperature, 
corresponding to a thermal broadening of the Fermi function of the order of the 
gap, a small peak at $E \approx \Delta$ emerges. This peak vanishes at 
low temperature when the broadening of the Fermi function is much smaller than 
the gap. The focus of this work is the systematic investigation of features 
due 
to this extra thermal channel, both from the theoretical and experimental point 
of view. In the following we distinguish between \emph{standard} 
resonance lines, which are also present at low temperatures, and \emph{thermal} 
lines due to the presence of the extra thermal peak.

\begin{figure}[ht]
\includegraphics[width=\columnwidth]{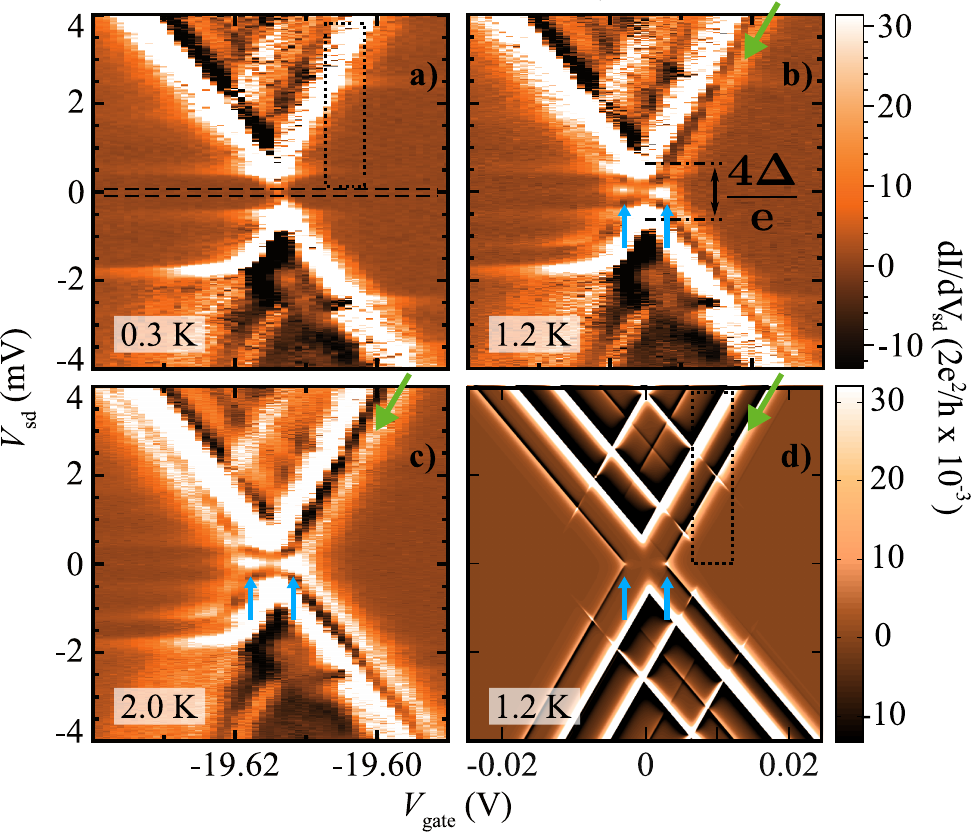}
\caption{%
(Color online) Differential conductance  $\text{d}I(\vsd,\vg) / \text{d}\vsd$ as 
a function of source-drain voltage \vsd\ and gate voltage \vg\ measured at (a) 
$T = \SI{0.3}{\kelvin}$, (b) $T=\SI{1.2}{\kelvin}$, (c) \mbox{$T = 
\SI{2.0}{\kelvin}$}, and (d) corresponding transport calculation at \mbox{$T = 
\SI{1.2}{\kelvin}$}. One of the additional lines emerging at high temperature 
is marked by a diagonal green arrow. Around zero bias two conductance peaks are 
clearly visible (vertical blue arrows). The dotted rectangles in (a) and (d) as 
well as the horizontal lines in (a) frame regions used to extract the line 
plots in Figs.~\ref{Fig:Figure4}(a) and (c). The maximum of the $\didv$
scale was set to \mbox{$0.031\times 2e^2/h$} to increase the contrast of the 
thermally induced lines.
}
\label{Fig:Figure2}
\end{figure}
Fig.~\ref{Fig:Figure2}(a)-(c) displays detailed measurements of the 
differential conductance at increasing temperatures, close to the charge 
degeneracy point marked by the black rectangle in Fig.~\ref{Fig:Figure1}(b).
\footnote{An apparent discrepancy in gate voltage range between 
Fig.~\ref{Fig:Figure1}(b) and Fig.~\ref{Fig:Figure2}(a)-(c) is due to long-time 
scale uniform drift of all Coulomb blockade features. The conductance peaks 
have been re-identified from overview measurements.}
The comparison of Fig.~\ref{Fig:Figure2}(a) and Figs.~\ref{Fig:Figure2}(b) and 
(c) gives direct evidence that at temperatures above $T \simeq 
\SI{300}{\milli\kelvin}$ additional transition lines parallel to 
the sequential tunneling lines emerge within the region of Coulomb blockade, 
see e.g.\ the green arrow in Fig.~\ref{Fig:Figure2}(b)-(d). These lines are 
separated from the sequential tunneling lines by a characteristic region of 
negative differential conductance (NDC, dark). As can be seen in 
Fig.~\ref{Fig:Figure2}(d), the additional lines and the NDC 
regions are reproduced by our transport calculations described in detail below, 
which account for sequential tunneling processes of thermally excited 
quasiparticles. At the intersection of such lines we obtain two zero bias 
conductance peaks indicated by blue arrows and separated by \makebox{$\delta 
V_g = 2 \left|\Delta\right|/e \alpha_{\text{{g}}}$}, with $\alpha_{\text{{g}}}$ 
as the gate coupling factor.

{\it Theoretical model--} 
Our calculations are based on a master equation approach for the reduced 
density matrix (RDM) to lowest order in the tunneling to the leads, including 
only quasiparticle tunneling.\cite{Pfaller2013} The theory is 
generalized here 
to include also the shell and orbital degrees of freedom of the CNT.
Specifically, the quantum dot is modeled by the Hamiltonian
\begin{equation}\label{eq:hop_cnt}
 \Hop{CNT}= \sum_{\alpha \sigma} \epsilon_{\alpha\sigma}
\dddag{\alpha \sigma} \dan{\alpha \sigma} + \frac{U}{2} \operatorname{\hat N}
\big(  \operatorname{\hat N} -1
\big),
\end{equation}  
where \makebox{$\alpha = (s,\tau)$} is a collective quantum number accounting 
for longitudinal $(s)$ and orbital $(\tau)$ degrees of freedom, respectively, 
and $\sigma$ 
labels the spin. 
 \footnote {The term longitudinal mode is referring to the
energy quantization associated to the finite length of the carbon nanotube. 
The  Hamiltonian in Eq.~(\ref{eq:hop_cnt}) includes both spin
orbit interaction and KK' mixing terms. Since it is written in the
diagonal basis, the quantum number $\tau$ is accounting for the ``orbital''
degrees of freedom, which are linear combinations of states in the \mbox{$KK'$}
basis. \cite{Flensberg2010}}
Finally, we employ a constant interaction 
model for the Coulomb repulsion on the tube with strength $U$. Including two 
longitudinal modes, \makebox{$s=1,2$}, and accounting for the two 
orbital degrees of freedom, \makebox{$\tau \in \{a,b\}$}, of the CNTs,  
\makebox{$\epsilon_{\alpha \sigma}$} represents four energy levels with 
energies \makebox{$\epsilon_0$}, \makebox{$\epsilon_0+\delta$}, 
\makebox{$\epsilon_0+\Delta\epsilon$}, and \makebox{$\epsilon_0 + 
\Delta\epsilon + \delta$}. The characteristic fourfold degeneracy of the carbon 
nanotube spectrum is assumed to be lifted by \makebox{$\delta = \sqrt{\DSO^2 + 
\DKK^2}$} originating from spin orbit splitting \DSO\ and 
valley-mixing \DKK.\cite{Flensberg2010} 

The size of the experimentally measured Coulomb diamonds and the positions of 
the excited state lines in the stability diagrams are consistent with the 
assumption that the transitions occur between states with \makebox{$(4n+3)$} 
and \makebox{$(4n+4)$} electrons. They are correctly reproduced in our model 
with  $\delta=1.3\un{meV}$, a spacing between the longitudinal modes 
\makebox{$\Delta\epsilon =1.55\,\delta$}, and \makebox{$U= 15\un{meV}$}. The
gate voltage is assumed to linearly shift the single particle 
energy levels 
\makebox{$\epsilon_{\alpha\sigma} \rightarrow \epsilon_{\alpha\sigma} +  
\alpha_{\text{g}} e V_g$}. At finite bias voltage the electrochemical 
potentials in 
the source and drain electrodes are \makebox{$\mu_{\text{\tiny{S/D}}} = \mu_0 
\pm \alpha_{\text{\tiny{S/D}}} eV_b $}, where 
\makebox{$\alpha_{\text{\tiny{S}}} = \alpha_{\text{\tiny{sd}}}$}
and \makebox{$\alpha_{\text{\tiny{D}}} =1- \alpha_{\text{\tiny{sd}}}$} account 
for the asymmetric bias drop at the source and drain contact, respectively.
From our simulations, we find an effective back gate coupling 
\makebox{$\alpha_{\text{{g}}} = 0.1$} and an asymmetric bias drop 
\makebox{$\alpha_{\text{\tiny{sd}}} = 0.4$}.

\begin{figure}
\includegraphics[width=\columnwidth]{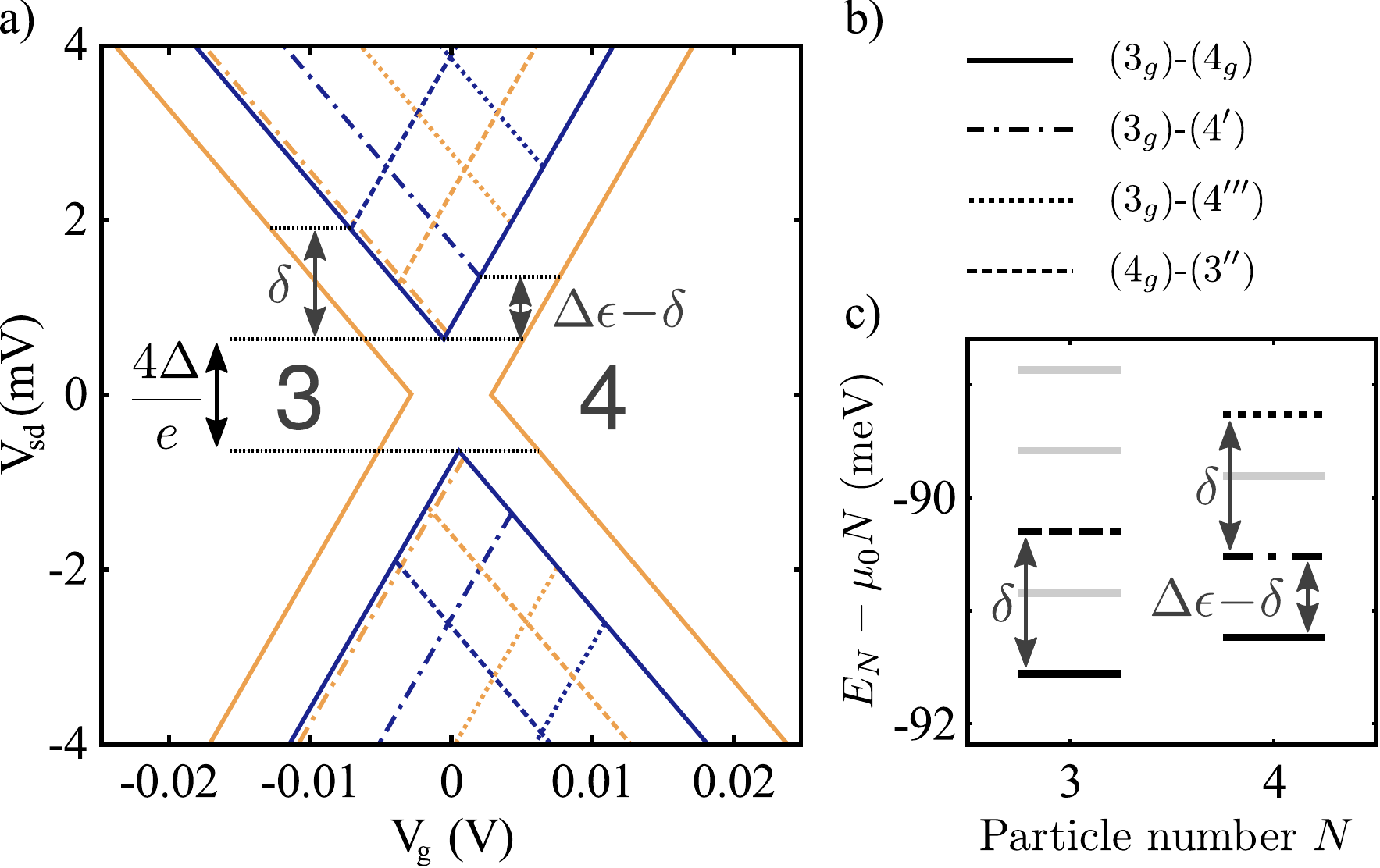}
\caption{ (Color online)
(a) Expected position of the differential conductance lines of the stability 
diagram  in  Fig.~\ref{Fig:Figure2}(d). The solid blue (dark gray) lines 
correspond to standard transitions between the \mbox{$(4n+3)$}- and 
\mbox{$(4n+4)$}-electron ground states, denoted ($3_g$) and ($4_g$), 
respectively. The solid orange (light gray) lines are caused by thermally 
activated transport channels. Lines from standard transitions involving an 
excited state are depicted as broken blue lines, the associated thermal replica 
in orange with the same line style.
(b) Legend associating transition lines to transitions between 
states.
We denoted the first, second, and third 
excited state of electron number $N$ by ($N'$), ($N''$), and ($N'''$), 
respectively.  (c) Many-body spectrum of the \mbox{($2n+3$)} and
\mbox{($2n+4$)} 
electron subspace as observed in transport, for \mbox{$\alpha_g eV_g = |\Delta| 
$}. 
Here the 3-particle groundstate energy is $E_3^g -3\mu_0 = -2 \delta -6U +3
|\Delta|$.
The distances between the energy levels $\Delta E$ (cf. 
Eqs.~(\ref{eq:transport_cond_s}) and (\ref{eq:transport_cond_th})), marked by 
arrows, are used to extract $\Delta\epsilon$ and $\delta$ from the 
measurements. Transitions involving the levels marked light gray are not
experimentally observed.
}
\label{Fig:Figure3}
\end{figure}
The expected positions of the differential conductance lines of the stability 
diagrams are displayed in Fig.~\ref{Fig:Figure3}(a). The solid blue lines show 
the $(4n+3)$ electron ground state to $(4n+4)$ electron ground state transition 
\makebox{$(3_g)$-$(4_g)$},
the broken blue lines are instead transition lines between a ground state and 
an excited state of the neighbouring particle number, see 
Fig.~\ref{Fig:Figure3}(b). Each of the possible standard transition lines is 
accompanied by an associated thermal line (in orange, same line style) due to 
thermally activated quasiparticles. We set the zero of the gate voltage 
at the charge degeneracy point. The position of the blue transition lines is 
then dictated by the standard sequential tunneling 
requirements,\cite{Pfaller2013}
\begin{equation}\label{eq:transport_cond_s}
 eV_{\text{sd}} = \frac{1}{\alpha_{\text{\tiny{S/D}} } } \bigg(   
 \pm\alpha_g eV_g+ \Delta E   + |\Delta|
 \bigg),
\end{equation} 
for source lines ($+$) and  drain ($-$) lines. Here, \mbox{$\Delta E$} is the 
energy difference between an excited state and a ground state with the same 
particle number in the many-body spectrum of Fig.~\ref{Fig:Figure3}(c). In the 
case of a source (drain) transition $\Delta E$ is calculated in the $N$
\mbox{($N+1$)} particle subspace. For a ground state to ground state 
transition, \mbox{$\Delta E = 0$} in Eq.~(\ref{eq:transport_cond_s}).

The conditions for the occurrence of an orange thermal line are 
\begin{equation}\label{eq:transport_cond_th}
 eV_{\text{sd}} = \frac{1}{\alpha_{\text{\tiny{S/D}} } } \bigg(   
 \pm \alpha_g eV_g + \Delta E   - |\Delta|
 \bigg).
\end{equation}
Thus, each replica runs parallel to the diamond edge at a distance 
\makebox{$2|\Delta|/\alpha_{\text{\tiny{S/D}}}$} from the standard line 
associated to it.

\begin{figure}
\includegraphics[width=\columnwidth]{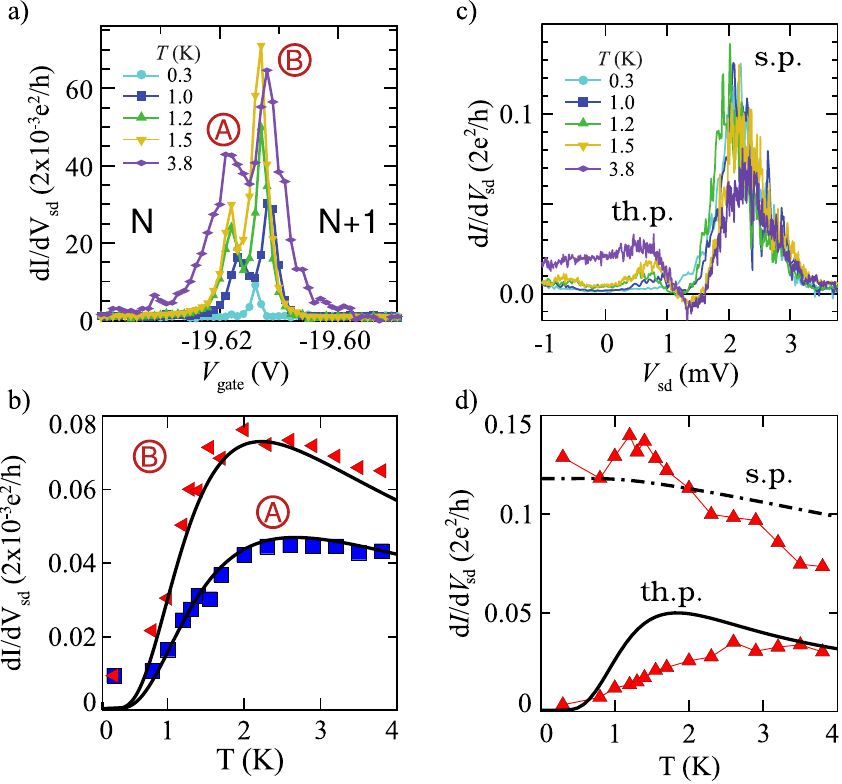}
\caption{(Color online) 
(a) Gate voltage dependence of the low bias conductance at different
temperatures. Each trace is an average over the bias voltage region marked in 
Fig.~\ref{Fig:Figure2}(a). (b) Temperature dependence of the conductance maxima 
A (squares) and B (triangles) in Fig.~\ref{Fig:Figure4}(a), together with a
model calculation according to Eq.~(\ref{eq:conductance}) (lines).
 (c) Bias traces of the differential conductance, 
taken within the rectangular area in Fig.~\ref{Fig:Figure2}(a), for different 
temperatures (see text). Two peaks
due to standard (s.p.) and thermal (th.p.) processes are observed. (d) 
Temperature dependence of the maximum of the differential conductance peaks of
Fig.~\ref{Fig:Figure4}(c). The solid and the 
dash-dotted line result from corresponding model 
calculations using a corresponding average. Our second order theory is
overestimating the peak height of the standard peak. Hence the curve was
multiplied by $0.28$ for a
better qualitative comparison.
} 
\label{Fig:Figure4}
\end{figure}
{\it Low bias conductance-- } Fig.~\ref{Fig:Figure4}(a) shows the gate voltage 
dependence of the low bias differential conductance for increasing temperature.
Each trace is an average of several measurements taken at small but finite 
bias values symmetrically located around $\vsd = 0$ and corresponding to the 
area between the dashed horizontal lines in Fig.~\ref{Fig:Figure2}(a). Note 
that due to the existence of a superconducting energy gap, no current would be 
expected in this bias voltage range. Two clearly distinguishable peaks are 
observed. They result from the  zero-bias crossing of the 
thermally induced transition lines. Due to their thermal nature, they decrease 
for decreasing temperature. At $T=\SI{0.3}{\kelvin}$ the double peak is absent. 
A single peak observed at approximately the position of the charge degeneracy 
point may be due to higher order processes not captured by the theory discussed 
below. 

In Fig.~\ref{Fig:Figure4}(b) the maximal conductance measured at the two peaks
denoted 
by A and B in Fig.~\ref{Fig:Figure4}(a) is plotted as a function of the
temperature (squares and triangles, respectively). The observed behaviour is
well reproduced by an analytic expression for the linear conductance
 derived 
around the $N$ to \mbox{$(N+1)$} charge degeneracy point (solid lines). By
taking into account the ground state energy levels of the relevant $N$ and
$N+1$-particles subspace, we find
\begin{equation}\label{eq:conductance}
\begin{split}
 \frac{dI}{dV_{sd}}\bigg|_{{V_{sd}=0}}   &= \frac{e^2}{2}\frac{\Gamma}{k_B
T} \operatorname{Re}\bigg ( \cosh\bigg( \frac{\Delta E_g +i\gamma}{2 k_B T}\bigg)
\bigg)^{-2}\,\\
&\times D(\Delta E_g ) \big( \rho_N + \rho_{N+1} \big),
\end{split}
\end{equation}
 with the BCS density of states  $D(E)$, and the occupation probability of the
$N$-particle ground state $\rho_N$. The energy difference  $\Delta E_g$  between
the two ground states scales linearly with \vg\ and equals zero at the charge 
degeneracy point.
Here,  $\gamma$  is a phenomenological Dynes parameter\cite{Dynes1978} 
related to a finite lifetime of the quasiparticles in the superconducting
leads. The Dynes parameter is introduced to renormalize the  BCS
density of states and therefore   leads to a
broadening of the conductance peaks. 
Eq.~(\ref{eq:conductance}) predicts the existence of two maxima of the
conductance located symmetrically around $\Delta E_g = 0$. The values of the
maxima as a function of the temperature correspond to the two solid lines in
Fig.~\ref{Fig:Figure4}(b).

 We notice a good agreement between 
experiment and theory at temperatures above \mbox{$T\sim \SI{1}{\kelvin}$}.
From the fit we extract the coupling parameter \makebox{$\Gamma=
0.093\un{meV}$} and the Dynes parameter
\makebox{$\gamma=0.015\un{meV}$}. 
The temperature dependence 
of the conductance peaks can be divided into two regimes. At low temperature, 
i.e., in the thermal activation regime, the ``$\cosh^{-2}$'' term of 
Eq.~(\ref{eq:conductance}) dominates, leading to a steep increase of the peaks 
with temperature. In the high temperature regime, we see the typical $1/T$ 
decay known from standard sequential tunneling processes.\cite{Beenakker1991}

Of particular interest is the ratio of the conductance maxima of the left and 
the right peak. Since Eq.~(\ref{eq:conductance}) is symmetric around
\mbox{$\Delta E_g = 0$}, we find that the ratio is equal to
\begin{equation}\label{eq:ratio}
\frac{dI/dV_{sd}^A}{dI/dV_{sd}^B} =
 \frac{(\rho_N + \rho_{N+1})_A}{(\rho_N + \rho_{N+1})_B} =
 \frac{d_{N+1} + d_N\,e^{-\Delta E_g/k_B T} }{d_{N} + d_{N+1}\,e^{-\Delta
E_g/k_B T}} , 
\end{equation} 
where canonical expressions were used for the occupation 
probabilities $\rho_N$ and $\rho_{N+1}$, and $d_N$ ($d_{N+1}$) denotes the 
degeneracy of the $N$ and $(N+1)$ particles ground state. Thus, it is possible 
to directly probe the degeneracy of the two ground states using 
Eq.~(\ref{eq:ratio}). Fig.~\ref{Fig:Figure4}(a) and (b) show that the 
conductance at point (B) is larger than at point (A), leading to the conclusion 
that the $N$-particles ground state has a larger degeneracy than the 
$N+1$-particles ground state. This confirms the assumption that the data are 
measured around a $(4n+3)$-$(4n+4)$ type charge degeneracy point, as also 
supported by the correspondence between the theory and experiment of excited 
state transition lines (see Fig.~\ref{Fig:Figure2}).

According to our model, the four fold degeneracy of the CNT is broken and the 
degeneracy of a ground state with odd number of particles, due to time reversal 
symmetry, equals $d_{2N+1} = 2$. Taking the ratio of the measured peak height 
of the two thermally induced conductance peaks provides, to our knowledge, a 
new method to determine the degeneracy of the ground state of multi-electron 
quantum dot single electron transistors.

{\it Finite bias conductance--}
The behavior of thermal and standard transitions at finite bias is depicted in
Fig.~\ref{Fig:Figure4}(c), which shows  $\didv(\vsd)$ traces at various 
temperatures. These traces result from an average taken over the voltage range 
marked by the dotted box in Fig.~\ref{Fig:Figure2}(a).\footnote{In order to not 
introduce an artificial broadening in the average, since the conductance peaks 
lie at different \vsd\ for different values of \vg, the curves were shifted 
to compensate for that offset. The reference with respect to which all other 
curves were shifted was always the curve closest to the degeneracy point. 
Repeating that procedure for different temperatures yields  
Fig.~\ref{Fig:Figure4}(c).} We observe two peaks which evolve in opposite ways 
at increasing temperature: the standard peak (s.p.) at higher \vsd\ decreases  
as expected from standard sequential tunneling.\cite{Beenakker1991} The second 
one at lower $\vsd$ increases and hence confirms thermally assisted 
quasiparticle tunneling (thermal peak, th.p.). A characteristic dip evolving 
into NDC is also clearly observed in the line traces in 
Fig.~\ref{Fig:Figure4}(c). 

In Fig.~\ref{Fig:Figure4}(d) the extracted temperature dependence of both the 
thermally activated and the standard sequential tunneling peak is depicted 
(triangles). Similar to the data analysis of the experiments, also the 
theoretical curves for the peak height were calculated via averaging over the 
voltage range marked with the dotted rectangle in Fig.~\ref{Fig:Figure2}(d).
Our perturbative theory is overestimating the height of the standard
peak. Hence, the theoretical curve (dash-dotted line) was
multiplied 
by $0.28$ to allow a better comparison with experimental data. A decrease of
the 
peak is observed with increasing temperature.  The calculation for the  
thermally activated peak (solid black line) is in good agreement with 
experiments; it shows a similar temperature dependence as the conductance peaks 
in Fig.~\ref{Fig:Figure4}(b). 

Preliminary calculations show that a renormalization of the lowest order theory 
taking into account also charge fluctuations in the framework of a dressed 
second order theory\cite{Kern2013} can reproduce the broadening 
(linewidth) of the resonance peaks and gives the correct ratio between the peak 
height of the thermally induced and the standard sequential tunneling peak. 
This study will be the subject of an upcoming publication. 

{\it Conclusions-- }
We demonstrate thermally activated quasiparticle transport in a carbon nanotube 
quantum dot with superconducting contacts. Our theoretical analysis shows that 
the new lines in the otherwise blockaded regions of the stability diagram 
appear already in the sequential tunneling regime. The splitting of the 
thermally induced conductance peaks at low bias can be used to probe 
the degeneracy of the ground states, and provides a particularly useful method 
to determine charge configurations from transport characteristics. 

The authors would like to thank Kicheon Kang for insightful discussions. We 
acknowledge funding from the Deutsche Forschungsgemeinschaft (SFB 631 TP A11,  
GRK 1570, Emmy Noether project Hu 1808-1) and from the EU FP7 Project SE2ND.

\bibliography{paper,manual}

\begin{thebibliography}{29}%
\makeatletter
\providecommand \@ifxundefined [1]{%
 \@ifx{#1\undefined}
}%
\providecommand \@ifnum [1]{%
 \ifnum #1\expandafter \@firstoftwo
 \else \expandafter \@secondoftwo
 \fi
}%
\providecommand \@ifx [1]{%
 \ifx #1\expandafter \@firstoftwo
 \else \expandafter \@secondoftwo
 \fi
}%
\providecommand \natexlab [1]{#1}%
\providecommand \enquote  [1]{``#1''}%
\providecommand \bibnamefont  [1]{#1}%
\providecommand \bibfnamefont [1]{#1}%
\providecommand \citenamefont [1]{#1}%
\providecommand \href@noop [0]{\@secondoftwo}%
\providecommand \href [0]{\begingroup \@sanitize@url \@href}%
\providecommand \@href[1]{\@@startlink{#1}\@@href}%
\providecommand \@@href[1]{\endgroup#1\@@endlink}%
\providecommand \@sanitize@url [0]{\catcode `\\12\catcode `\$12\catcode
  `\&12\catcode `\#12\catcode `\^12\catcode `\_12\catcode `\%12\relax}%
\providecommand \@@startlink[1]{}%
\providecommand \@@endlink[0]{}%
\providecommand \url  [0]{\begingroup\@sanitize@url \@url }%
\providecommand \@url [1]{\endgroup\@href {#1}{\urlprefix }}%
\providecommand \urlprefix  [0]{URL }%
\providecommand \Eprint [0]{\href }%
\providecommand \doibase [0]{http://dx.doi.org/}%
\providecommand \selectlanguage [0]{\@gobble}%
\providecommand \bibinfo  [0]{\@secondoftwo}%
\providecommand \bibfield  [0]{\@secondoftwo}%
\providecommand \translation [1]{[#1]}%
\providecommand \BibitemOpen [0]{}%
\providecommand \bibitemStop [0]{}%
\providecommand \bibitemNoStop [0]{.\EOS\space}%
\providecommand \EOS [0]{\spacefactor3000\relax}%
\providecommand \BibitemShut  [1]{\csname bibitem#1\endcsname}%
\let\auto@bib@innerbib\@empty
\bibitem [{\citenamefont {Bockrath}\ \emph {et~al.}(1997)\citenamefont
  {Bockrath}, \citenamefont {Cobden}, \citenamefont {McEuen}, \citenamefont
  {Chopra}, \citenamefont {Zettl}, \citenamefont {Thess},\ and\ \citenamefont
  {Smalley}}]{Bockrath1997a}%
  \BibitemOpen
  \bibfield  {author} {\bibinfo {author} {\bibfnamefont {M.}~\bibnamefont
  {Bockrath}}, \bibinfo {author} {\bibfnamefont {D.~H.}\ \bibnamefont
  {Cobden}}, \bibinfo {author} {\bibfnamefont {P.~L.}\ \bibnamefont {McEuen}},
  \bibinfo {author} {\bibfnamefont {N.~G.}\ \bibnamefont {Chopra}}, \bibinfo
  {author} {\bibfnamefont {A.}~\bibnamefont {Zettl}}, \bibinfo {author}
  {\bibfnamefont {A.}~\bibnamefont {Thess}}, \ and\ \bibinfo {author}
  {\bibfnamefont {R.~E.}\ \bibnamefont {Smalley}},\ }\href@noop {} {\bibfield
  {journal} {\bibinfo  {journal} {Science}\ }\textbf {\bibinfo {volume}
  {275}},\ \bibinfo {pages} {1922} (\bibinfo {year} {1997})}\BibitemShut
  {NoStop}%
\bibitem [{\citenamefont {Jensen}\ \emph {et~al.}(2005)\citenamefont {Jensen},
  \citenamefont {Hauptmann}, \citenamefont {Nyg\aa{}rd},\ and\ \citenamefont
  {Lindelof}}]{prb-jensen}%
  \BibitemOpen
  \bibfield  {author} {\bibinfo {author} {\bibfnamefont {A.}~\bibnamefont
  {Jensen}}, \bibinfo {author} {\bibfnamefont {J.~R.}\ \bibnamefont
  {Hauptmann}}, \bibinfo {author} {\bibfnamefont {J.}~\bibnamefont
  {Nyg\aa{}rd}}, \ and\ \bibinfo {author} {\bibfnamefont {P.~E.}\ \bibnamefont
  {Lindelof}},\ }\href@noop {} {\bibfield  {journal} {\bibinfo  {journal}
  {Phys. Rev. B}\ }\textbf {\bibinfo {volume} {72}},\ \bibinfo {pages} {035419}
  (\bibinfo {year} {2005})}\BibitemShut {NoStop}%
\bibitem [{\citenamefont {Kasumov}\ \emph {et~al.}(2003)\citenamefont
  {Kasumov}, \citenamefont {Kociak}, \citenamefont {Ferrier}, \citenamefont
  {Deblock}, \citenamefont {Gu\'eron}, \citenamefont {Reulet}, \citenamefont
  {Khodos}, \citenamefont {St\'ephan},\ and\ \citenamefont
  {Bouchiat}}]{prb-kasumov}%
  \BibitemOpen
  \bibfield  {author} {\bibinfo {author} {\bibfnamefont {A.}~\bibnamefont
  {Kasumov}}, \bibinfo {author} {\bibfnamefont {M.}~\bibnamefont {Kociak}},
  \bibinfo {author} {\bibfnamefont {M.}~\bibnamefont {Ferrier}}, \bibinfo
  {author} {\bibfnamefont {R.}~\bibnamefont {Deblock}}, \bibinfo {author}
  {\bibfnamefont {S.}~\bibnamefont {Gu\'eron}}, \bibinfo {author}
  {\bibfnamefont {B.}~\bibnamefont {Reulet}}, \bibinfo {author} {\bibfnamefont
  {I.}~\bibnamefont {Khodos}}, \bibinfo {author} {\bibfnamefont
  {O.}~\bibnamefont {St\'ephan}}, \ and\ \bibinfo {author} {\bibfnamefont
  {H.}~\bibnamefont {Bouchiat}},\ }\href@noop {} {\bibfield  {journal}
  {\bibinfo  {journal} {Phys. Rev. B}\ }\textbf {\bibinfo {volume} {68}},\
  \bibinfo {pages} {214521} (\bibinfo {year} {2003})}\BibitemShut {NoStop}%
\bibitem [{\citenamefont {Grove-Rasmussen}\ \emph {et~al.}(2009)\citenamefont
  {Grove-Rasmussen}, \citenamefont {J\o{}rgensen}, \citenamefont {Andersen},
  \citenamefont {Paaske}, \citenamefont {Jespersen}, \citenamefont
  {Nyg\aa{}rd}, \citenamefont {Flensberg},\ and\ \citenamefont
  {Lindelof}}]{prb-groverasmussen}%
  \BibitemOpen
  \bibfield  {author} {\bibinfo {author} {\bibfnamefont {K.}~\bibnamefont
  {Grove-Rasmussen}}, \bibinfo {author} {\bibfnamefont {H.~I.}\ \bibnamefont
  {J\o{}rgensen}}, \bibinfo {author} {\bibfnamefont {B.~M.}\ \bibnamefont
  {Andersen}}, \bibinfo {author} {\bibfnamefont {J.}~\bibnamefont {Paaske}},
  \bibinfo {author} {\bibfnamefont {T.~S.}\ \bibnamefont {Jespersen}}, \bibinfo
  {author} {\bibfnamefont {J.}~\bibnamefont {Nyg\aa{}rd}}, \bibinfo {author}
  {\bibfnamefont {K.}~\bibnamefont {Flensberg}}, \ and\ \bibinfo {author}
  {\bibfnamefont {P.~E.}\ \bibnamefont {Lindelof}},\ }\href@noop {} {\bibfield
  {journal} {\bibinfo  {journal} {Phys. Rev. B}\ }\textbf {\bibinfo {volume}
  {79}},\ \bibinfo {pages} {134518} (\bibinfo {year} {2009})}\BibitemShut
  {NoStop}%
\bibitem [{\citenamefont {Cleuziou}\ \emph {et~al.}(2006)\citenamefont
  {Cleuziou}, \citenamefont {Wernsdorfer}, \citenamefont {Bouchiat},
  \citenamefont {Ondarcuhu},\ and\ \citenamefont {Monthioux}}]{nnano-cleuziou}%
  \BibitemOpen
  \bibfield  {author} {\bibinfo {author} {\bibfnamefont {J.-P.}\ \bibnamefont
  {Cleuziou}}, \bibinfo {author} {\bibfnamefont {W.}~\bibnamefont
  {Wernsdorfer}}, \bibinfo {author} {\bibfnamefont {V.}~\bibnamefont
  {Bouchiat}}, \bibinfo {author} {\bibfnamefont {T.}~\bibnamefont {Ondarcuhu}},
  \ and\ \bibinfo {author} {\bibfnamefont {M.}~\bibnamefont {Monthioux}},\
  }\href@noop {} {\bibfield  {journal} {\bibinfo  {journal} {Nat. Nanotech.}\
  }\textbf {\bibinfo {volume} {1}},\ \bibinfo {pages} {53} (\bibinfo {year}
  {2006})}\BibitemShut {NoStop}%
\bibitem [{\citenamefont {Jarillo-Herrero}\ \emph {et~al.}(2006)\citenamefont
  {Jarillo-Herrero}, \citenamefont {van Dam},\ and\ \citenamefont
  {Kouwenhoven}}]{JarilloHerrero2006}%
  \BibitemOpen
  \bibfield  {author} {\bibinfo {author} {\bibfnamefont {P.}~\bibnamefont
  {Jarillo-Herrero}}, \bibinfo {author} {\bibfnamefont {J.~A.}\ \bibnamefont
  {van Dam}}, \ and\ \bibinfo {author} {\bibfnamefont {L.~P.}\ \bibnamefont
  {Kouwenhoven}},\ }\href@noop {} {\bibfield  {journal} {\bibinfo  {journal}
  {Nature}\ }\textbf {\bibinfo {volume} {439}},\ \bibinfo {pages} {953}
  (\bibinfo {year} {2006})}\BibitemShut {NoStop}%
\bibitem [{\citenamefont {Pallecchi}\ \emph {et~al.}(2008)\citenamefont
  {Pallecchi}, \citenamefont {Gaass}, \citenamefont {Ryndyk},\ and\
  \citenamefont {Strunk}}]{Pallecchi2008}%
  \BibitemOpen
  \bibfield  {author} {\bibinfo {author} {\bibfnamefont {E.}~\bibnamefont
  {Pallecchi}}, \bibinfo {author} {\bibfnamefont {M.}~\bibnamefont {Gaass}},
  \bibinfo {author} {\bibfnamefont {D.~A.}\ \bibnamefont {Ryndyk}}, \ and\
  \bibinfo {author} {\bibfnamefont {C.}~\bibnamefont {Strunk}},\ }\href@noop {}
  {\bibfield  {journal} {\bibinfo  {journal} {Appl. Phys. Lett.}\ }\textbf
  {\bibinfo {volume} {93}},\ \bibinfo {pages} {072501} (\bibinfo {year}
  {2008})}\BibitemShut {NoStop}%
\bibitem [{\citenamefont {Dirks}\ \emph {et~al.}(2011)\citenamefont {Dirks},
  \citenamefont {Hughes}, \citenamefont {Lal}, \citenamefont {Uchoa},
  \citenamefont {Chen}, \citenamefont {Chialvo}, \citenamefont {Goldbart},\
  and\ \citenamefont {Mason}}]{nphys-dirks}%
  \BibitemOpen
  \bibfield  {author} {\bibinfo {author} {\bibfnamefont {T.}~\bibnamefont
  {Dirks}}, \bibinfo {author} {\bibfnamefont {T.~L.}\ \bibnamefont {Hughes}},
  \bibinfo {author} {\bibfnamefont {S.}~\bibnamefont {Lal}}, \bibinfo {author}
  {\bibfnamefont {B.}~\bibnamefont {Uchoa}}, \bibinfo {author} {\bibfnamefont
  {Y.-F.}\ \bibnamefont {Chen}}, \bibinfo {author} {\bibfnamefont
  {C.}~\bibnamefont {Chialvo}}, \bibinfo {author} {\bibfnamefont {P.~M.}\
  \bibnamefont {Goldbart}}, \ and\ \bibinfo {author} {\bibfnamefont
  {N.}~\bibnamefont {Mason}},\ }\href@noop {} {\bibfield  {journal} {\bibinfo
  {journal} {Nat. Phys.}\ }\textbf {\bibinfo {volume} {7}},\ \bibinfo {pages}
  {386} (\bibinfo {year} {2011})}\BibitemShut {NoStop}%
\bibitem [{\citenamefont {Pillet}\ \emph {et~al.}(2010)\citenamefont {Pillet},
  \citenamefont {Quay}, \citenamefont {Morfin}, \citenamefont {Bena},
  \citenamefont {Yeyati},\ and\ \citenamefont {Joyez}}]{nphys-pillet}%
  \BibitemOpen
  \bibfield  {author} {\bibinfo {author} {\bibfnamefont {J.-D.}\ \bibnamefont
  {Pillet}}, \bibinfo {author} {\bibfnamefont {C.~H.~L.}\ \bibnamefont {Quay}},
  \bibinfo {author} {\bibfnamefont {P.}~\bibnamefont {Morfin}}, \bibinfo
  {author} {\bibfnamefont {C.}~\bibnamefont {Bena}}, \bibinfo {author}
  {\bibfnamefont {A.~L.}\ \bibnamefont {Yeyati}}, \ and\ \bibinfo {author}
  {\bibfnamefont {P.}~\bibnamefont {Joyez}},\ }\href@noop {} {\bibfield
  {journal} {\bibinfo  {journal} {Nat. Phys.}\ }\textbf {\bibinfo {volume}
  {6}},\ \bibinfo {pages} {965} (\bibinfo {year} {2010})}\BibitemShut {NoStop}%
\bibitem [{\citenamefont {Kim}\ \emph {et~al.}(2013)\citenamefont {Kim},
  \citenamefont {Ahn}, \citenamefont {Kim}, \citenamefont {Choi}, \citenamefont
  {Bae}, \citenamefont {Kang}, \citenamefont {Lim}, \citenamefont {L\'{o}pez},\
  and\ \citenamefont {Kim}}]{Kim2013}%
  \BibitemOpen
  \bibfield  {author} {\bibinfo {author} {\bibfnamefont {B.-K.}\ \bibnamefont
  {Kim}}, \bibinfo {author} {\bibfnamefont {Y.-H.}\ \bibnamefont {Ahn}},
  \bibinfo {author} {\bibfnamefont {J.-J.}\ \bibnamefont {Kim}}, \bibinfo
  {author} {\bibfnamefont {M.-S.}\ \bibnamefont {Choi}}, \bibinfo {author}
  {\bibfnamefont {M.-H.}\ \bibnamefont {Bae}}, \bibinfo {author} {\bibfnamefont
  {K.}~\bibnamefont {Kang}}, \bibinfo {author} {\bibfnamefont {J.~S.}\
  \bibnamefont {Lim}}, \bibinfo {author} {\bibfnamefont {R.}~\bibnamefont
  {L\'{o}pez}}, \ and\ \bibinfo {author} {\bibfnamefont {N.}~\bibnamefont
  {Kim}},\ }\href@noop {} {\bibfield  {journal} {\bibinfo  {journal} {Phys.
  Rev. Lett.}\ }\textbf {\bibinfo {volume} {110}},\ \bibinfo {pages} {076803}
  (\bibinfo {year} {2013})}\BibitemShut {NoStop}%
\bibitem [{\citenamefont {Pillet}\ \emph {et~al.}(2013)\citenamefont {Pillet},
  \citenamefont {Joyez}, \citenamefont {\v{Z}itko},\ and\ \citenamefont
  {Goffman}}]{Pillet2013}%
  \BibitemOpen
  \bibfield  {author} {\bibinfo {author} {\bibfnamefont {J.-D.}\ \bibnamefont
  {Pillet}}, \bibinfo {author} {\bibfnamefont {P.}~\bibnamefont {Joyez}},
  \bibinfo {author} {\bibfnamefont {R.}~\bibnamefont {\v{Z}itko}}, \ and\
  \bibinfo {author} {\bibfnamefont {M.~F.}\ \bibnamefont {Goffman}},\
  }\href@noop {} {\bibfield  {journal} {\bibinfo  {journal} {Phys. Rev. B}\
  }\textbf {\bibinfo {volume} {88}},\ \bibinfo {pages} {045101} (\bibinfo
  {year} {2013})}\BibitemShut {NoStop}%
\bibitem [{\citenamefont {Kumar}\ \emph {et~al.}(2014)\citenamefont {Kumar},
  \citenamefont {Gaim}, \citenamefont {Steininger}, \citenamefont
  {Levy~Yeyati}, \citenamefont {Mart\'{i}n-Rodero}, \citenamefont
  {H\"{u}ttel},\ and\ \citenamefont {Strunk}}]{Kumar2014}%
  \BibitemOpen
  \bibfield  {author} {\bibinfo {author} {\bibfnamefont {A.}~\bibnamefont
  {Kumar}}, \bibinfo {author} {\bibfnamefont {M.}~\bibnamefont {Gaim}},
  \bibinfo {author} {\bibfnamefont {D.}~\bibnamefont {Steininger}}, \bibinfo
  {author} {\bibfnamefont {A.}~\bibnamefont {Levy~Yeyati}}, \bibinfo {author}
  {\bibfnamefont {A.}~\bibnamefont {Mart\'{i}n-Rodero}}, \bibinfo {author}
  {\bibfnamefont {A.~K.}\ \bibnamefont {H\"{u}ttel}}, \ and\ \bibinfo {author}
  {\bibfnamefont {C.}~\bibnamefont {Strunk}},\ }\href@noop {} {\bibfield
  {journal} {\bibinfo  {journal} {Phys. Rev. B}\ }\textbf {\bibinfo {volume}
  {89}},\ \bibinfo {pages} {075428} (\bibinfo {year} {2014})}\BibitemShut
  {NoStop}%
\bibitem [{\citenamefont {Pfaller}\ \emph {et~al.}(2013)\citenamefont
  {Pfaller}, \citenamefont {Donarini},\ and\ \citenamefont
  {Grifoni}}]{Pfaller2013}%
  \BibitemOpen
  \bibfield  {author} {\bibinfo {author} {\bibfnamefont {S.}~\bibnamefont
  {Pfaller}}, \bibinfo {author} {\bibfnamefont {A.}~\bibnamefont {Donarini}}, \
  and\ \bibinfo {author} {\bibfnamefont {M.}~\bibnamefont {Grifoni}},\
  }\href@noop {} {\bibfield  {journal} {\bibinfo  {journal} {Phys. Rev. B}\
  }\textbf {\bibinfo {volume} {87}},\ \bibinfo {pages} {155439} (\bibinfo
  {year} {2013})}\BibitemShut {NoStop}%
\bibitem [{\citenamefont {Kong}\ \emph {et~al.}(1998)\citenamefont {Kong},
  \citenamefont {Soh}, \citenamefont {Cassell}, \citenamefont {Quate},\ and\
  \citenamefont {Dai}}]{Kong1998}%
  \BibitemOpen
  \bibfield  {author} {\bibinfo {author} {\bibfnamefont {J.}~\bibnamefont
  {Kong}}, \bibinfo {author} {\bibfnamefont {H.}~\bibnamefont {Soh}}, \bibinfo
  {author} {\bibfnamefont {A.}~\bibnamefont {Cassell}}, \bibinfo {author}
  {\bibfnamefont {C.}~\bibnamefont {Quate}}, \ and\ \bibinfo {author}
  {\bibfnamefont {H.}~\bibnamefont {Dai}},\ }\href@noop {} {\bibfield
  {journal} {\bibinfo  {journal} {Nature}\ }\textbf {\bibinfo {volume} {395}},\
  \bibinfo {pages} {878} (\bibinfo {year} {1998})}\BibitemShut {NoStop}%
\bibitem [{\citenamefont {Martinis}\ and\ \citenamefont
  {Kautz}(1989)}]{Martinis1989}%
  \BibitemOpen
  \bibfield  {author} {\bibinfo {author} {\bibfnamefont {J.~M.}\ \bibnamefont
  {Martinis}}\ and\ \bibinfo {author} {\bibfnamefont {R.~L.}\ \bibnamefont
  {Kautz}},\ }\href@noop {} {\bibfield  {journal} {\bibinfo  {journal} {Phys.
  Rev. Lett.}\ }\textbf {\bibinfo {volume} {63}},\ \bibinfo {pages} {1507}
  (\bibinfo {year} {1989})}\BibitemShut {NoStop}%
\bibitem [{\citenamefont {{De Franceschi}}\ \emph {et~al.}(2001)\citenamefont
  {{De Franceschi}}, \citenamefont {Sasaki}, \citenamefont {Elzerman},
  \citenamefont {van~der Wiel}, \citenamefont {Tarucha},\ and\ \citenamefont
  {Kouwenhoven}}]{DeFranceschi2001}%
  \BibitemOpen
  \bibfield  {author} {\bibinfo {author} {\bibfnamefont {S.}~\bibnamefont {{De
  Franceschi}}}, \bibinfo {author} {\bibfnamefont {S.}~\bibnamefont {Sasaki}},
  \bibinfo {author} {\bibfnamefont {J.~M.}\ \bibnamefont {Elzerman}}, \bibinfo
  {author} {\bibfnamefont {W.~G.}\ \bibnamefont {van~der Wiel}}, \bibinfo
  {author} {\bibfnamefont {S.}~\bibnamefont {Tarucha}}, \ and\ \bibinfo
  {author} {\bibfnamefont {L.~P.}\ \bibnamefont {Kouwenhoven}},\ }\href@noop {}
  {\bibfield  {journal} {\bibinfo  {journal} {Phys. Rev. Lett.}\ }\textbf
  {\bibinfo {volume} {86}},\ \bibinfo {pages} {878} (\bibinfo {year}
  {2001})}\BibitemShut {NoStop}%
\bibitem [{\citenamefont {Holm}\ \emph {et~al.}(2008)\citenamefont {Holm},
  \citenamefont {J{\o}rgensen}, \citenamefont {Grove-Rasmussen}, \citenamefont
  {Paaske}, \citenamefont {Flensberg},\ and\ \citenamefont
  {Lindelof}}]{Holm2008}%
  \BibitemOpen
  \bibfield  {author} {\bibinfo {author} {\bibfnamefont {J.~V.}\ \bibnamefont
  {Holm}}, \bibinfo {author} {\bibfnamefont {H.~I.}\ \bibnamefont
  {J{\o}rgensen}}, \bibinfo {author} {\bibfnamefont {K.}~\bibnamefont
  {Grove-Rasmussen}}, \bibinfo {author} {\bibfnamefont {J.}~\bibnamefont
  {Paaske}}, \bibinfo {author} {\bibfnamefont {K.}~\bibnamefont {Flensberg}}, \
  and\ \bibinfo {author} {\bibfnamefont {P.~E.}\ \bibnamefont {Lindelof}},\
  }\href@noop {} {\bibfield  {journal} {\bibinfo  {journal} {Phys. Rev. B}\
  }\textbf {\bibinfo {volume} {77}},\ \bibinfo {pages} {161406} (\bibinfo
  {year} {2008})}\BibitemShut {NoStop}%
\bibitem [{Note1()}]{Note1}%
  \BibitemOpen
  \bibinfo {note} {An evaluation of the elastic cotunneling lines in Fig.~\ref
  {Fig:Figure1}(b), not within the scope of our lowest-order theory, results in
  a slightly reduced value $\Delta \sim \SI {250}{\micro \electronvolt
  }$.}\BibitemShut {Stop}%
\bibitem [{\citenamefont {Hulm}\ \emph {et~al.}(1972)\citenamefont {Hulm},
  \citenamefont {Jones}, \citenamefont {Hein},\ and\ \citenamefont
  {Gibson}}]{Hulm1972}%
  \BibitemOpen
  \bibfield  {author} {\bibinfo {author} {\bibfnamefont {J.~K.}\ \bibnamefont
  {Hulm}}, \bibinfo {author} {\bibfnamefont {C.~K.}\ \bibnamefont {Jones}},
  \bibinfo {author} {\bibfnamefont {R.~A.}\ \bibnamefont {Hein}}, \ and\
  \bibinfo {author} {\bibfnamefont {J.~W.}\ \bibnamefont {Gibson}},\
  }\href@noop {} {\bibfield  {journal} {\bibinfo  {journal} {Journal of Low
  Temperature Physics}\ }\textbf {\bibinfo {volume} {7}},\ \bibinfo {pages}
  {291} (\bibinfo {year} {1972})}\BibitemShut {NoStop}%
\bibitem [{\citenamefont {{De Franceschi}}\ \emph {et~al.}(2010)\citenamefont
  {{De Franceschi}}, \citenamefont {Kouwenhoven}, \citenamefont
  {Sch\"{o}nenberger},\ and\ \citenamefont {Wernsdorfer}}]{DeFranceschi2010}%
  \BibitemOpen
  \bibfield  {author} {\bibinfo {author} {\bibfnamefont {S.}~\bibnamefont {{De
  Franceschi}}}, \bibinfo {author} {\bibfnamefont {L.}~\bibnamefont
  {Kouwenhoven}}, \bibinfo {author} {\bibfnamefont {C.}~\bibnamefont
  {Sch\"{o}nenberger}}, \ and\ \bibinfo {author} {\bibfnamefont
  {W.}~\bibnamefont {Wernsdorfer}},\ }\href@noop {} {\bibfield  {journal}
  {\bibinfo  {journal} {Nat. Nanotech.}\ }\textbf {\bibinfo {volume} {5}},\
  \bibinfo {pages} {703} (\bibinfo {year} {2010})}\BibitemShut {NoStop}%
\bibitem [{\citenamefont {Buitelaar}\ \emph {et~al.}(2003)\citenamefont
  {Buitelaar}, \citenamefont {Belzig}, \citenamefont {Nussbaumer},
  \citenamefont {Babi\'{c}}, \citenamefont {Bruder},\ and\ \citenamefont
  {Sch\"{o}nenberger}}]{Buitelaar2003}%
  \BibitemOpen
  \bibfield  {author} {\bibinfo {author} {\bibfnamefont {M.}~\bibnamefont
  {Buitelaar}}, \bibinfo {author} {\bibfnamefont {W.}~\bibnamefont {Belzig}},
  \bibinfo {author} {\bibfnamefont {T.}~\bibnamefont {Nussbaumer}}, \bibinfo
  {author} {\bibfnamefont {B.}~\bibnamefont {Babi\'{c}}}, \bibinfo {author}
  {\bibfnamefont {C.}~\bibnamefont {Bruder}}, \ and\ \bibinfo {author}
  {\bibfnamefont {C.}~\bibnamefont {Sch\"{o}nenberger}},\ }\href@noop {}
  {\bibfield  {journal} {\bibinfo  {journal} {Phys. Rev. Lett.}\ }\textbf
  {\bibinfo {volume} {91}},\ \bibinfo {pages} {057005} (\bibinfo {year}
  {2003})}\BibitemShut {NoStop}%
\bibitem [{\citenamefont {Goldhaber-Gordon}\ \emph {et~al.}(1998)\citenamefont
  {Goldhaber-Gordon}, \citenamefont {G\"{o}res}, \citenamefont {Kastner},
  \citenamefont {Shtrikman}, \citenamefont {Mahalu},\ and\ \citenamefont
  {Meirav}}]{GoldhaberPRL1998}%
  \BibitemOpen
  \bibfield  {author} {\bibinfo {author} {\bibfnamefont {D.}~\bibnamefont
  {Goldhaber-Gordon}}, \bibinfo {author} {\bibfnamefont {J.}~\bibnamefont
  {G\"{o}res}}, \bibinfo {author} {\bibfnamefont {M.~A.}\ \bibnamefont
  {Kastner}}, \bibinfo {author} {\bibfnamefont {H.}~\bibnamefont {Shtrikman}},
  \bibinfo {author} {\bibfnamefont {D.}~\bibnamefont {Mahalu}}, \ and\ \bibinfo
  {author} {\bibfnamefont {U.}~\bibnamefont {Meirav}},\ }\href@noop {}
  {\bibfield  {journal} {\bibinfo  {journal} {Phys. Rev. Lett.}\ }\textbf
  {\bibinfo {volume} {81}},\ \bibinfo {pages} {5225} (\bibinfo {year}
  {1998})}\BibitemShut {NoStop}%
\bibitem [{Note2()}]{Note2}%
  \BibitemOpen
  \bibinfo {note} {An apparent discrepancy in gate voltage range between
  Fig.~\ref {Fig:Figure1}(b) and Fig.~\ref {Fig:Figure2}(a)-(c) is due to
  long-time scale uniform drift of all Coulomb blockade features. The
  conductance peaks have been re-identified from overview
  measurements.}\BibitemShut {Stop}%
\bibitem [{Note3()}]{Note3}%
  \BibitemOpen
  \bibinfo {note} {The term longitudinal mode is referring to the energy
  quantization associated to the finite length of the carbon nanotube. The
  Hamiltonian in Eq.~(\ref {eq:hop_cnt}) includes both spin orbit interaction
  and KK' mixing terms. Since it is written in the diagonal basis, the quantum
  number $\tau $ is accounting for the ``orbital'' degrees of freedom, which
  are linear combinations of states in the \unhbox \voidb@x \hbox {$KK'$}
  basis. \cite {Flensberg2010}}\BibitemShut {NoStop}%
\bibitem [{\citenamefont {Flensberg}\ and\ \citenamefont
  {Marcus}(2010)}]{Flensberg2010}%
  \BibitemOpen
  \bibfield  {author} {\bibinfo {author} {\bibfnamefont {K.}~\bibnamefont
  {Flensberg}}\ and\ \bibinfo {author} {\bibfnamefont {C.~M.}\ \bibnamefont
  {Marcus}},\ }\href@noop {} {\bibfield  {journal} {\bibinfo  {journal} {Phys.
  Rev. B}\ }\textbf {\bibinfo {volume} {81}},\ \bibinfo {pages} {195418}
  (\bibinfo {year} {2010})}\BibitemShut {NoStop}%
\bibitem [{\citenamefont {Dynes}\ \emph {et~al.}(1978)\citenamefont {Dynes},
  \citenamefont {Narayanamurti},\ and\ \citenamefont {Garno}}]{Dynes1978}%
  \BibitemOpen
  \bibfield  {author} {\bibinfo {author} {\bibfnamefont {R.~C.}\ \bibnamefont
  {Dynes}}, \bibinfo {author} {\bibfnamefont {V.}~\bibnamefont
  {Narayanamurti}}, \ and\ \bibinfo {author} {\bibfnamefont {J.~P.}\
  \bibnamefont {Garno}},\ }\href@noop {} {\bibfield  {journal} {\bibinfo
  {journal} {Phys. Rev. Lett.}\ }\textbf {\bibinfo {volume} {41}},\ \bibinfo
  {pages} {1509} (\bibinfo {year} {1978})}\BibitemShut {NoStop}%
\bibitem [{\citenamefont {Beenakker}(1991)}]{Beenakker1991}%
  \BibitemOpen
  \bibfield  {author} {\bibinfo {author} {\bibfnamefont {C.~W.~J.}\
  \bibnamefont {Beenakker}},\ }\href@noop {} {\bibfield  {journal} {\bibinfo
  {journal} {Phys. Rev. B}\ }\textbf {\bibinfo {volume} {44}},\ \bibinfo
  {pages} {1646} (\bibinfo {year} {1991})}\BibitemShut {NoStop}%
\bibitem [{Note4()}]{Note4}%
  \BibitemOpen
  \bibinfo {note} {In order to not introduce an artificial broadening in the
  average, since the conductance peaks lie at different \protect \ensuremath
  {V_{\protect \text {sd}}}\ for different values of \protect \ensuremath
  {V_{\protect \text {gate}}}, the curves were shifted to compensate for that
  offset. The reference with respect to which all other curves were shifted was
  always the curve closest to the degeneracy point. Repeating that procedure
  for different temperatures yields Fig.~\ref {Fig:Figure4}(c).}\BibitemShut
  {Stop}%
\bibitem [{\citenamefont {Kern}\ and\ \citenamefont
  {Grifoni}(2013)}]{Kern2013}%
  \BibitemOpen
  \bibfield  {author} {\bibinfo {author} {\bibfnamefont {J.}~\bibnamefont
  {Kern}}\ and\ \bibinfo {author} {\bibfnamefont {M.}~\bibnamefont {Grifoni}},\
  }\href@noop {} {\bibfield  {journal} {\bibinfo  {journal} {Eur. Phys. J. B}\
  }\textbf {\bibinfo {volume} {86}},\ \bibinfo {pages} {384} (\bibinfo {year}
  {2013})}\BibitemShut {NoStop}%
\end{thebibliography}%

\end{document}